\documentclass[aip,twocolumn,preprintnumbers,superscriptaddress,showpacs]{revtex4}
\usepackage{amssymb}
\usepackage{graphicx}
\usepackage{dcolumn}
\usepackage{bm}
\usepackage{multirow}

\begin{document}

\title{Cyrstal field splitting and optical band gap of hexagonal LuFeO$_3$ films}

\author{Wenbin Wang}
\affiliation{Department of Physics, University of Tennessee, Knoxville, TN 37996, USA}
\affiliation{Materials Science and Technology Division, Oak Ridge National Laboratory, Oak Ridge, TN 37831, USA}

\author{Hongwei Wang}
\affiliation{Department of Physics and Institute for Computational Molecular Science,
Temple University, Philadelphia, PA 19122, USA}

\author{Xiaoying Xu}
\affiliation{Materials Science and Technology Division, Oak Ridge National Laboratory, Oak Ridge, TN 37831, USA}

\author{Leyi Zhu}
\affiliation{Materials Science Division, Argonne National Laboratory, Argonne, IL 60439, USA}

\author{Lixin He}
\affiliation{Key Laboratory of Quantum Information,
University of Science and Technology of China, Hefei, Anhui 230026, China}

\author{Elizabeth Wills}
\affiliation{Department of Physics, Bryn Mawr College, Bryn Mawr, PA 19010, USA}

\author{Xuemei Cheng}
\affiliation{Department of Physics, Bryn Mawr College, Bryn Mawr, PA 19010, USA}

\author{David J. Keavney}
\affiliation{Advanced Photon Source, Argonne National Laboratory, Argonne, IL 60439, USA}

\author{Jian Shen}
\affiliation{Department of Physics, University of Tennessee, Knoxville, TN 37996, USA}
\affiliation{Department of Physics, Fudan University, Shanghai 200433, China}

\author{Xifan Wu$^*$}
\affiliation{Department of Physics and Institute for Computational Molecular Science,
Temple University, Philadelphia, PA 19122, USA}

\author{Xiaoshan Xu$^*$}
\affiliation{Materials Science and Technology Division, Oak Ridge National Laboratory, Oak Ridge, TN 37831, USA}

\date{\today }

\begin{abstract}
 Hexagonal LuFeO$_3$ films have been studied using x-ray absorption and optical spectroscopy. 
 The crystal splittings of Fe$^{3+}$ are extracted as $E_{e'}-E_{e''}$=0.7 eV and $E_{a_1'}-E_{e'}$=0.9 eV and a 2.0 eV optical band gap is determined assuming a direct gap. 
 First-principles calculations confirm the experiments that the relative energies of crystal field splitting states do follow $E_{a_1'}>E_{e'}>E_{e''}$ with slightly underestimated values and a band gap of 1.35 eV.
\end{abstract}


\maketitle

\clearpage

  Hexagonal mangnites (RMnO$_3$, R=Sc, Y, Ho-Lu) belong to multiferroic materials that simultaneously exhibit more than one type of ferroic order, advantageous over other existing materials in terms of new types of information processing and storage.\cite{Khomskii2005,Spaldin2010}
  The robust ferroelectricity in RMnO$_3$ ($T_C$$\sim$1000 K, $P$$\sim$ 5 $\mu C/cm^2$) originates from the instability of the MnO$_5$ trigonal-bipyramid rotation that renders soft zone-center modes involving the vertical displacement of R atoms, breaking the inversion symmetry.\cite{Fennie2005,Chupis1982,Fujimura1996}
 The ordering of magnetic moments in RMnO$_3$ sets in at T$_N$$\sim$100 K, below which the interesting magnetoelectric coupling has been demonstrated.\cite{Muoz2000,Lottermoser2004}

 On the other hand, hexagonal RFeO$_3$ (h-RFeO$_3$) are not stable in free standing bulk form. 
 They can be stabilized by quenching a levitated melt in an aerodynamic levitation furnace or in solvothermal reactions.\cite{Kuribayashi2008, Kumar2008, Kumar2009, Magome2010}
 In addition, h-LuFeO$_3$ has also been stabilized in films using pulsed laser deposition (PLD) and metal-organic chemical vapor deposition on yttrium stabilized zirconium oxides (YSZ) and Al$_2$O$_3$.\cite{Bossak2004, Akbashev2011, Jeong2012, Iida2012} 
 The structural characterization indicates that h-LuFeO$_3$ is isomorphic to RMnO$_3$ with space group $P6_3$cm (185).
 The similar crystallographic structure of h-RFeO$_3$ and RMnO$_3$ suggest that h-RFeO$_3$ is also ferroelectric.\cite{Bossak2004,Magome2010,Jeong2012} 
 The larger spin on Fe$^{3+}$ compared with Mn$^{3+}$ in principle generates stronger spin-spin interactions, allowing for higher magnetic ordering temperature, which has been suggested by recent experiments.\cite{Akbashev2011,Wang2012,Jeong2012} 
 In stark contrast to the large amount of work devoted to RMnO$_3$, detailed studies on the h-RFeO$_3$, and especially of the electronic structures, are still lacking.\cite{McCarthy1973,Chupis1982,Thomas2010,Pavlov2012}

 In this paper, we investigate the electronic properties of h-LuFeO$_3$ films using x-ray absorption and optical spectroscopy. 
 The analysis of the x-ray absorption spectra (XAS) using crystal field theory reveal a splitting of the Fe $3d$ levels significantly higher than that of Fe$^{3+}$ in Lu$_2$FeO$_4$, indicating stronger Fe-O interactions. 
 The extracted optical band gap from optical absorption spectra is 2.0 $\pm$ 0.1 eV, somewhat smaller than that of perovskite ferrites.\cite{Xu2009}
 The experimental findings have been confirmed by our electronic structure calculations.
 
\begin{figure}[b]
\centerline{
\includegraphics[width = 3.0 in]{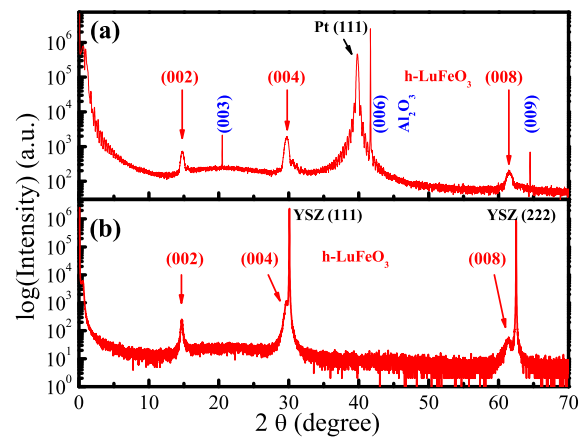}}
\caption{ (Color online)
The x-ray diffraction spectra of h-LuFeO$_3$ films grown on (a) Al$_2$O$_3$ substrates with Pt buffer layer and (b) YSZ substrates.
}
 \label{fig-xrd}
\end{figure}

 The XAS was studied on 50 nm h-LuFeO$_3$ films grown on Al$_2$O$_3$ substrates using PLD with 30 nm Pt buffer layer (Fig.\ref{fig-xrd}(a)) to avoid charging effect. 
 The XAS was taken at beam line 4-ID-C at the Advanced Photon Source using polarized synchrotron x-rays. 
 A 20 nm thick h-LuFeO$_3$ film was grown on a YSZ substrate using PLD (Fig.\ref{fig-xrd}(b)) for optical spectroscopy measurements. 
 Part of the substrate was covered by a mask at growth so it can be used as a reference in the optical transmittance measurements.
 Optical spectra were collected in transmittance mode using a Varian Cary 5000 spectrometer.


\begin{figure}[t]
\centerline{
\includegraphics[width = 3.0 in]{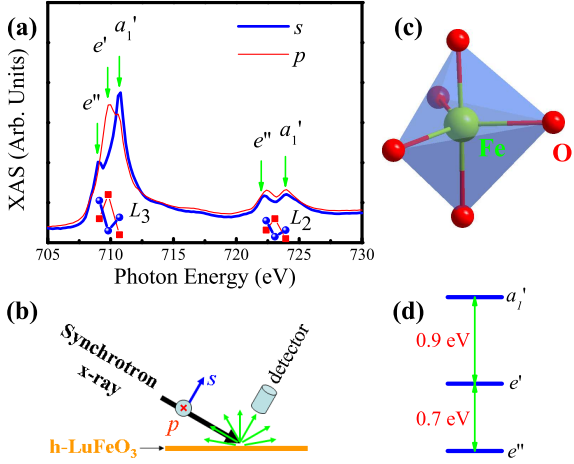}}
\caption{(Color online) 
 Crystal field splittings of Fe site (300 K). 
 (a) The x-ray absorption spectra corresponding to Fe $2p$ to Fe $3d$ excitations. 
 The symbols are the calculated matrix elements from the initial to the final on-electron states (see text). 
 (b) Schematics of the experimental setup. 
 (c) Schematics of the local environment of Fe sites.
 (d) The crystal splittings extracted from the XAS spectra.
 }
\label{fig-xas}
\end{figure}

 Figure \ref{fig-xas}(a) shows the XAS corresponding to transitions from a Fe $2p^63d^5$ to a Fe $2p^53d^6$ multiplet.
 In the spectra, two groups of peaks separated by approximately 12 eV can be distinguished. 
 For each group, fine structures depending on the polarization can be recognized: two well-separated peaks (709.1 and 710.7 eV) are observed for the \textit{s} polarization, while an additional intensity is observed for \textit{p} polarization as a peaks at 709.8 eV.

 These spectra details are determined by dipole and spin selection rules and a combination of effects from crystal-field, spin-orbit coupling, $d$-$p$ and $d$-$d$ interactions and Fe $3d$ O $2p$ hybridization.\cite{Laan1986,Stohr2006}
 In terms of one-electron energy, the Fe $2p$ states are split into $2p_{1/2}$ and $2p_{3/2}$ by the spin-orbit coupling, which has the energy scale of 15 eV, resulting in the two groups of excitations $L_2$ ($2p_{1/2}$ $\rightarrow$ $3d$) and $L_3$ ($2p_{3/2}$ $\rightarrow$ $3d$) in Fig. \ref{fig-xas}(a).\cite{Wadati2005,Stohr2006}
 For the Fe $3d$ states, the one-electron states are mainly split by crystal fields, which is on the order of one eV.\cite{Wadati2005,Stohr2006}
 Here the trigonal-bipyramid local environment of Fe gives rise to a symmetry that can be represented by the $D_{3h}$ point group as a good approximation, as shown in Fig. \ref{fig-xas}(c). 
 In this case, Fe $2p$ orbitals can be reduced to states corresponding to irreducible representations (IR) $e'(x,y)$ and $a_2''(z)$ while Fe $3d$ states split into IR $e''(xz,yz)$, $e'(xx-yy,xy)$ and $a_1'(zz)$.\cite{Cotton1990}
 A recent work on LuFe$_2$O$_4$ in which Fe$^{3+}$ also sit in a trigonal-bipyramid local environment, has shown that the energies of these crystal field states follow $E_{a_1'}>E_{e'}>E_{e''}$.\cite{Ko2009} 
 Similar results are also found for Mn$^{3+}$ in a $D_{3h}$ symmetry. \cite{Cho2007}

 According to Hund's rule, the ground states of Fe$^{3+}$ are a multiplet $^6A_1'$ for a $2p_{3/2}^42p_{1/2}^2 3d_{e''}^23d_{e'}^23d_{a_1'}^1$ electronic configuration. 
 In the ionic model that ignores hybridization between Fe $3d$ and O $2p$ state, the spin-allowed excited state multiplets and the corresponding one-electron state populations can be listed as shown in TABLE \ref{tab:selectionrules}.\cite{Laan1986,Note_p-d_splitting}
 Since the ground state multiplet has a symmetry of $A_1'$, the dipole-allowed excited states need to contain $E'$ or $A_2''$ to satisfy the dipole selection rules for a $D_{3h}$ symmetry.\cite{Cotton1990}
 The resulting dipole-allowed transitions are listed in TABLE \ref{tab:selectionrules}.
 It is clear that the photon with $z$ polarization can not excite an electron from Fe $2p$ to Fe $3d_{e'}$ state. 
 These selection rules are verified in the $L_3$ part of the XAS: for $s$ polarization, the intensity in the middle is much weaker, which suggests that the three peaks at 709.1, 709.8 and 710.7 eV are coming from the effect of crystal field.

 The transition probability in the XAS depends on the matrix elements $|<\psi_i|\hat{E}\cdot\vec{r}|\psi_f>|^2$, where the $\psi_i$ and $\psi_f$ are the initial and final one-electron states, $\hat{E}$ is the direction vector of the electric field, and $\vec{r}$ is the position vector. 
 As shown in Fig.\ref{fig-xas}(a) as symbols, the calculated matrix elements qualitatively agree with the dichroism for $L_3$ excitation. 
 The less obvious agreement for $L_2$ excitations is presumably due to a mixed energy splitting from crystal field and $p$-$d$ interactions.\cite{Stohr2006} 
 The peak positions allow for a rough determination of the crystal field splittings assuming similar $d$-$d$ interactions for different Fe $3d$ states: $E_{e'}-E_{e''}$=0.7 eV and $E_{a_1'}-E_{e'}$=0.9 eV, as also shown in Fig. \ref{fig-xas}(d). 
 These splittings are significantly larger than those in LuFe$_2$O$_4$ ($E_{e'}-E_{e''}$=0.4 eV and $E_{a_1'}-E_{e'}$=0.8 eV for Fe$^{3+}$ sites).\cite{Ko2009}
  The differences indicate a stronger Fe-O interactions, as also suggested by the different Fe-O bond length in h-LuFeO$_3$ and in LuFe$_2$O$_4$.\cite{Magome2010,Isobe1990}
 
\begin{table}[h]
\caption{The spin-allowed excited states and the dipole selection rules from the $^6A_1'(2p_{3/2}^42p_{1/2}^2 3d_{e''}^23d_{e'}^23d_{a_1'}^1)$ ground state with a linearly polarized photon.}
\label{tab:selectionrules}\centering
\begin{tabular}{c|c|c|c}
\hline\hline
$2p$ & $3d$ & IR & Allowed \\ \hline
\multirow{3}{*}{$2p_{3/2}^42p_{1/2}^1$} & $3d_{e''}^33d_{e'}^23d_{a_1'}^1$ & $A_1''$+$A_2''$+$E'$+$E''$ & $x,y,z$ \\ 
 & $3d_{e''}^23d_{e'}^33d_{a_1'}^1$ & $A_1''$+$A_2'$+$E'$+$E''$ & $x,y$  \\
 & $3d_{e''}^23d_{e'}^23d_{a_1'}^2$ & $E'$+$A_2''$ & $x,y,z$ \\  \hline
\multirow{3}{*}{$2p_{3/2}^32p_{1/2}^2$} & $3d_{e''}^32d_{e'}^23d_{a_1'}^1$ & $A_1''$+$A_2''$+$E'$+$E''$ & $x,y,z$ \\
 & $3d_{e''}^22d_{e'}^33d_{a_1'}^1$ & $A_1''$+$A_2'$+$E'$+$E''$ & $x,y$  \\ 
 & $3d_{e''}^22d_{e'}^23d_{a_1'}^2$ & $E'$+$A_2''$ & $x,y,z$  \\ \hline\hline
\end{tabular}
\end{table}


\begin{figure}[tbh]
\centerline{
\includegraphics[width = 3.0 in]{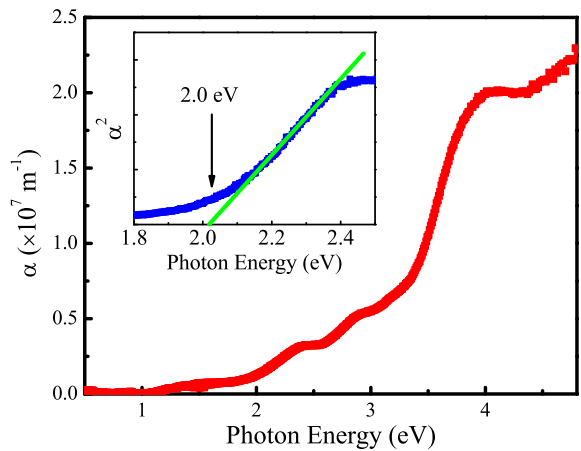}}
\caption{ (Color online)
Optical absorption coefficient $\alpha$ as a
function of photon energy. Inset: $\alpha^2$ as a function of photon energy which indicates an optical bandgap of 2.0 eV.
}
\label{fig-bandgap}
\end{figure}

 In order to further elucidate the electronic structure of the h-LuFeO$_3$, we measured optical absorption of the h-LuFeO$_3$ films. 
 The observed spectra (Fig.\ref{fig-bandgap}) shows three peak-like features at approximately 2.3, 2.9 and 3.9 eV, consistent with the recently reported optical properties of h-HoFeO$_3$ and h-ErFeO$_3$ films.\cite{Pavlov2012}
 Based upon the overall intensity $\sim 10^7$ cm$^{-1}$, these peaks correlate to dipole-allowed excitations. 
 Since the Fe$^{3+}$ has a $3d^5$ configuration, all the on-site excitations are spin forbidden. 
 Therefore, the peak at 2.3 eV is coming from charge transfer excitations.\cite{Xu2009} 
 A 2.0 $\pm$ 0.1 eV optical band gap was extracted using plots of $\alpha^2$ versus energy (Fig.\ref{fig-bandgap} inset), assuming a direct gap.

\begin{figure}[h]
\centerline{
\includegraphics[width = 3.0in]{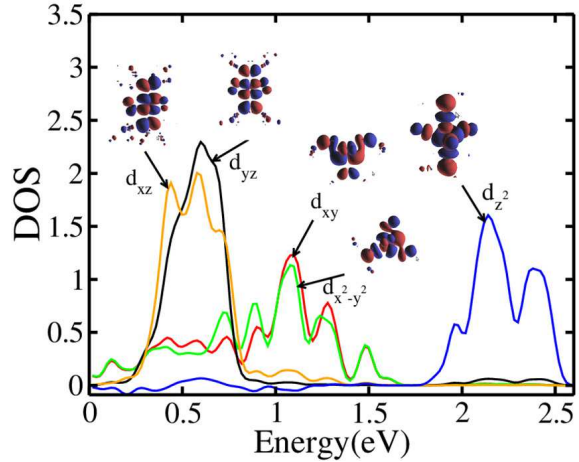}}
\caption{ (Color online) Projected density of states for conduction band of h-LuFeO$_3$ with theoretical ground state 
structure (space group $P6_{3}cm$) and MLWFs (by WIEN2k software package ~\cite{WIEN2k, WIEN2k_details}) with crystal field splitting states characters.
} 
\label{fig-dft}
\end{figure}

 First-principles electronic structure calculations can provide insightful picture of crystal field splitting. 
 We determined our projected density of states (PDOS) by the density functional theory (DFT) implemented in the Vienna ab initio simulations package (VASP)~\cite{kresse93,VASP_details}.
 We adopted the Perdew-Burke-Ernzerhof functional revised for solids (PBEsol)~\cite{perdew08} in which the spin-polarized generalized gradient approximation (GGA) is made in treating the exchange correlation effect of electrons.
 The resulting PDOS is presented in Fig.~\ref{fig-dft}. 
 One can clearly see that our theoretical results are consistent with the experimental data where the crystal field states follow $E_{a_1'}>E_{e'}>E_{e''}$.
 An unambiguous assignment of the crystal field states and energies can be further obtained by generating Maximally 
Localized Wannier Functions (MLWFs)~\cite{MLWF} based on the ground state electronic structure in the selected energy
window spanning all the crystal field states under consideration. 
 The resulting MLWFs for each crystal field state are shown in Fig.~\ref{fig-dft} and the resulting energies are $E_{e'}-E_{e''}=0.41~{\rm eV}$ and $E_{a_1'}-E_{e'}=0.81~{\rm eV}$, which are close to the experimental values of 0.7 and 0.9 eV respectively. 
 This qualitative agreement between experiment and theory is expected within the frame work of DFT. However we should be aware that a more proper treatment of electron-hole excitations by GW based Bethe-Salpeter method~\cite{Onida,XAS_Chen} can further improve the theoretical prediction.

To overcome the severely underestimated band gap due to the delocalization error arising from the
incomplete cancellation of the spurious self-interaction, we used the GGA+U with the effective U value 
(U$_{\rm eff}$=U-J) of 4.61 eV~\cite{WhangbL_2007}.
This gives a band gap of $E_{\rm g}=1.35~eV$ which still underestimates our experimental value. Again, a more proper treatment of self-energy by GW method will further bring the theoretical predictions closer to the experimental value.   

 In conclusion, we have studied h-LuFeO$_3$ films using x-ray and optical spectroscopy. 
 Clear dichroism observed in XAS is attributed to the effect of crystal field splitting which is found significantly larger than that of Fe$^{3+}$ in LuFe$_2$O$_4$, suggesting stronger Fe-O interaction. 
 A 2.0 eV optical band gap originated from charge transfer excitations is determined from the optical spectra. 
 This important information of electron structure, confirmed by DFT calculations will definitely benefit further studies of h-LuFeO$_3$.

 Research supported by the US DOE, BES MSED (X.S.X.), 
 and partially supported by the Chinese 973 Program Grant No. 2011CB921801 (J.S.), the US DOE BES Grant DE-SC0002136 (W.B.W.) and the NSF Grant No. 1053854 (X.M.C.).
 Use of the APS supported by the US DOE, BES Contract No. DE-AC02-06CH11357.

$^*$ To whom correspondence should be addressed: xiaoshan.xu@gatech.edu, xifanwu@temple.edu.


\begin{references}
\bibitem{VASP_details}
A 500 eV planewave cutoff and 6$\times$6$\times$3 $k$ points mesh are used.
\bibitem{WIEN2k_details}
The energy threshold separating the localized and nonlocalized electronic states is chosen to be 
-7Ry. The muffin-tin radii are 2.35 Bohr for Lu, 1.98 Bohr for Fe, and 1.65 Bohr for O. The criterion 
for the number of plane waves is chosen to be $R^{\rm min}_{\rm MT}*K^{\rm max} = 7$, and the number of
$k$ points is $8\times\times8\times2$ for A-type antiferromagnetic calculation.
\end{references}
\end{document}